\documentclass[conference]{IEEEtran}

\usepackage[normalem]{ulem}
\usepackage[T1]{fontenc}
\usepackage[utf8]{inputenc}
\usepackage{microtype}
\usepackage{amsmath,amssymb,amsfonts}
\usepackage{bm}
\usepackage{graphicx}
\usepackage{booktabs}
\usepackage{multirow}
\usepackage{xcolor}
\usepackage{enumitem}
\usepackage{array}
\usepackage{url}
\usepackage[caption=false,font=footnotesize]{subfig}
\usepackage{hyperref}
\usepackage{cleveref}

\hypersetup{
    colorlinks=true,
    linkcolor=blue,
    citecolor=blue,
    urlcolor=blue
}

\title{Fair Comparison of Scheduling Algorithms on Heterogeneous Edge Clusters: A Continuous Adaptive Benchmark }

\author{
    \IEEEauthorblockN{Zihang Wang\IEEEauthorrefmark{1}, 
                      Boris Sedlak\IEEEauthorrefmark{2},
                      Juan Luis Herrera\IEEEauthorrefmark{3}, and 
                      Schahram Dustdar\IEEEauthorrefmark{1}\IEEEauthorrefmark{4}}
    
    \IEEEauthorblockA{\IEEEauthorrefmark{1} TU Wien, Vienna, Austria}
    \IEEEauthorblockA{\IEEEauthorrefmark{2} Universitat Pompeu Fabra, Barcelona, Spain}
    \IEEEauthorblockA{\IEEEauthorrefmark{3} Universidad de Extremadura, Mérida, Spain}
    \IEEEauthorblockA{\IEEEauthorrefmark{4} ICREA, Barcelona, Spain}
    \IEEEauthorblockA{Corresponding author: zwang@dsg.tuwien.ac.at
    }
}
 
\begin{document}

\maketitle
\pagestyle{plain}

\begin{abstract}
Modern Artificial Intelligence (AI) workloads deployed across the heterogeneous tiers of an edge--cloud continuum must satisfy multi-dimensional Service Level Objectives (SLOs) over latency, throughput, and output quality. For each incoming task, the scheduler picks both a target node and a processing mode (e.g., full or reduced inference precision). We call this class of problems \emph{Continuous Multi-Mode Scheduling} (CMMS). Comparing CMMS algorithms fairly is difficult because prior studies typically evaluate each controller in its own stack, under a single workload, and without reporting per-decision overhead. To close these gaps, we present an open source benchmark platform that features (i) a unified controller interface, (ii) a closed-loop workload driver covering multiple workload patterns, and (iii) dual-metric SLO scoring that reports raw SLO (overall compliance) and steady-state SLO (compliance during stable operation) separately. Running six controllers across five cluster configurations and two load regimes (424 episodes), we find that controller rankings are strongly configuration-dependent: a deep reinforcement-learning winner under light workloads loses to a rule-based heuristic by nearly 29 percentage points once load intensifies, at roughly 500$\times$ the per-decision operational overhead. We further show that separating raw from steady-state SLOs exposes switching costs that a single aggregate score would otherwise conflate.
\end{abstract}

\section{Introduction}
Modern Artificial Intelligence (AI) workloads are increasingly deployed across the heterogeneous tiers of the computing continuum. The continuum spans edge devices, fog nodes, and accelerators with heterogeneous resources that are often required to sustain multi-dimensional Service Level Objectives (SLOs) over latency, throughput, and output quality~\cite{sedlak2024equilibrium,lapkovskis2025benchmarking}. This forces workload schedulers to make continuous runtime decisions: for each incoming task, dispatch it to a suitable node and optimize the execution (e.g., picking an AI model to process data). We call this class of problems \emph{Continuous Multi-Mode Scheduling} (CMMS). Common algorithms and solutions to CMMS problems can be fundamentally different, including rule-based heuristics~\cite{zhang2017live}, profile-guided optimization~\cite{gujarati2020clockwork}, deep reinforcement learning~\cite{mao2019learning}, and active inference~\cite{wang2025aif,sedlak2025adaptive}. Although this holds, a basic question remains hard to answer: \emph{Which class of controller makes the best continuous decisions to maximize SLOs under pressure, and at what cost?} 

To this end, we identify three gaps in how CMMS controllers are currently evaluated that make it difficult to fairly compare existing algorithms. First, the algorithm-specific testbeds conflate the algorithm with its test environment. Consequently, published numbers reflect the combination rather than the algorithm alone. For instance, VideoStorm~\cite{zhang2017live}, Chameleon~\cite{jiang2018chameleon}, and Pensieve~\cite{mao2017pensieve} each report results on their own custom testbed, which a third-party scheduler cannot reuse. Second, evaluating an algorithm within only a single scenario hides how different scheduling algorithms behave when the workload pattern changes, letting a  "single-scenario champion" be mistaken for a global optimum. Specifically, JCAB~\cite{zhang2020jcab} and Decima~\cite{mao2019learning} evaluate on a single workload pattern, leaving cross-scenario robustness unexamined. Third, the compute cost of the decision itself is rarely accounted for, though on Edge devices this directly consumes resources that otherwise could be dedicated for processing the workloads. Neither Pensieve~\cite{mao2017pensieve} nor Decima~\cite{mao2019learning} report per-decision latency alongside task-level SLO. Collectively, these gaps are particularly acute in \emph{Distributed Computing Continuum Systems} (DCCS), where heterogeneity of devices and tiers is the rule rather than the exception. In such settings, device heterogeneity and time-varying workloads can make single-testbed, single-scenario, overhead-blind evaluation particularly misleading.

To address these limitations, we present a fairness-first benchmark platform for CMMS in DCCS. The platform enables fair cross-algorithm evaluation through a unified controller interface, diverse workload scenarios that stress the system in different ways, and explicit accounting of per-decision overhead, all on a real heterogeneous continuum testbed under a conjunctive multi-dimensional SLO.

Using this benchmarking platform, we show that scheduler performance is strongly environment-dependent and that higher decision cost does not necessarily lead to better SLO compliance.

Thus, the main contributions of this work include:

\begin{itemize}
    \item A CMMS benchmark platform with a unified controller interface on a real heterogeneous edge cluster, enabling fair comparison across algorithms by reducing testbed-specific integration effects.

    \item A four-scenario evaluation methodology that exercises controllers under different operating conditions, reducing the risk of conclusions that depend on a single workload setting.

    \item A comparative evaluation of six controllers from three control paradigms, showing that controller behavior varies across scenarios and that per-decision overhead provides important context for interpreting SLO outcomes.
\end{itemize}

\section{Background}
Although many systems have been proposed for CMMS, fair comparison across them remains difficult. Existing work usually evaluates each scheduler in its own stack, under limited workloads, and often without accounting for decision overhead (\S\ref{sec:bench_challenges}). Benchmarks from related areas address parts of this problem, but not the full CMMS setting (\S\ref{sec:prior_benchmarks}).

\subsection{Why Fair Comparison Does Not Exist}
\label{sec:bench_challenges}

CMMS-style control has been tackled by a wide spectrum of schedulers, each developed inside its own evaluation environment. Video analytics systems couple profiling, configuration search, and execution into tightly integrated prototypes~\cite{zhang2017live,jiang2018chameleon}; adaptive bitrate controllers are trained and evaluated inside their own network simulator~\cite{mao2017pensieve}; cluster-job schedulers target a single datacenter workload trace~\cite{mao2019learning,zhang2020jcab}. More recently, active inference has been applied to edge AI service routing on a heterogeneous DCCS testbed~\cite{wang2025aif} and to adaptive stream processing under multi-dimensional SLOs~\cite{sedlak2025adaptive}, broadening the controller space beyond rule-based heuristics and reinforcement learning. In every case, however, the scheduling algorithm is distributed together with its own profiling or execution harness, making results hard to compare across papers. Most studies also consider only narrow workload conditions, leaving robustness under dynamic demand unclear. In addition, scheduler overhead is frequently omitted, even though on edge platforms a single decision can consume a noticeable fraction of a \(200\,\mathrm{ms}\) SLO budget. As a result, the literature still lacks a clean head-to-head comparison of CMMS algorithms across design families.

\subsection{Why Existing Benchmarks Do Not Help}
\label{sec:prior_benchmarks}

General-purpose benchmarks address related concerns but not CMMS directly. MLPerf~\cite{reddi2020mlperf,mattson2020mlperf} fixes reference models and workloads to compare inference backends, not scheduling algorithms that operate on top of those backends. DeathStarBench~\cite{gan2019deathstarbench} provides representative microservice workloads, but its focus is the infrastructure and service-mesh substrate rather than per-request scheduling decisions. SPEC elasticity benchmarks~\cite{herbst2015bungee} target IaaS auto-scaling, i.e.,\ adding or removing replicas in response to load; this differs structurally from continuous multi-mode dispatch, which changes the per-stream processing parameters at each control step. Consequently, none of these benchmarks provides a unified way to evaluate CMMS policies under dynamic workloads while accounting for controller overhead on heterogeneous edge hardware. This gap motivates the benchmark platform presented next.

\section{Platform Design}
\label{sec:design}

\begin{figure}[!t]
\centering
\includegraphics[width=\columnwidth,height=0.75\textheight,keepaspectratio]{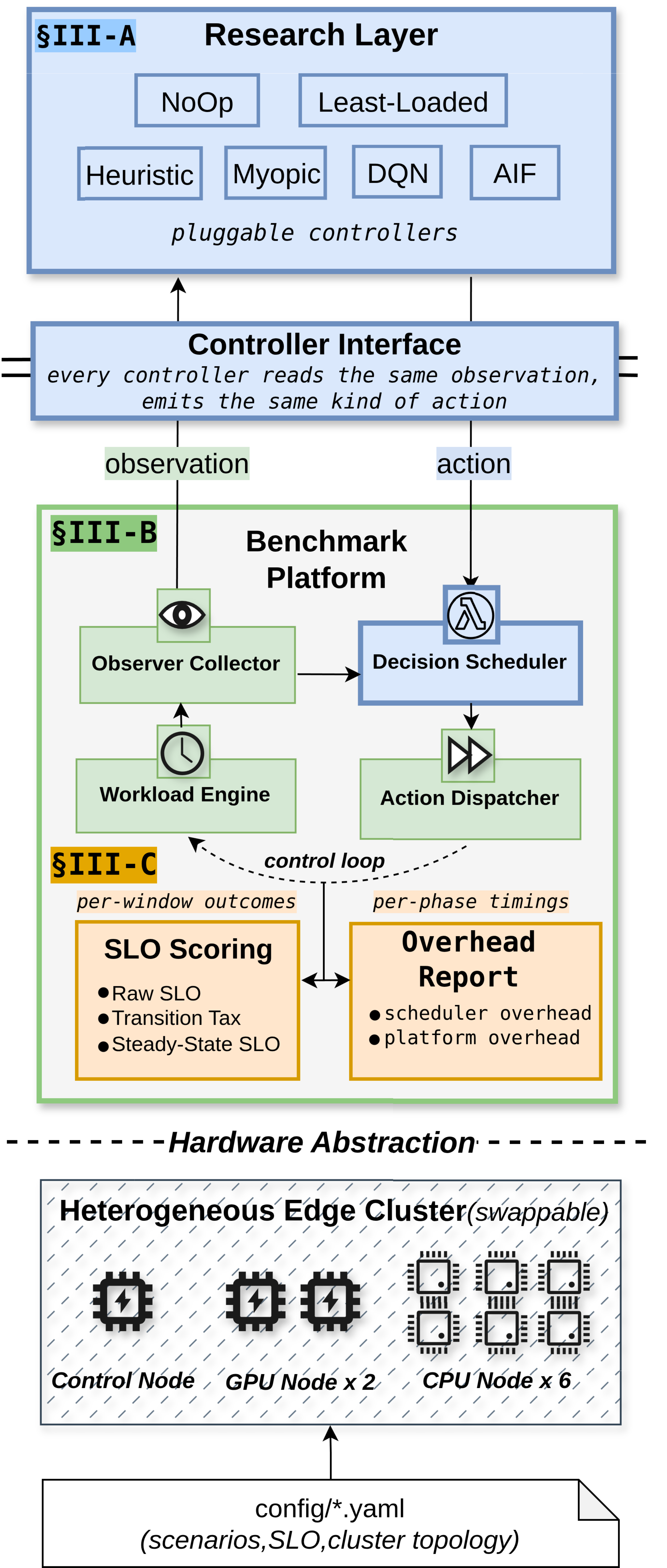}
\caption{Three-layer platform architecture. The fixed middle layer comprises the unified interface, closed-loop evaluation harness, and scoring stage.}
\label{fig:arch}
\end{figure}

The comparison of CMMS solutions remains constrained by testbed coupling, single-scenario masking, and invisible decision overhead. In response, our platform, as shown in Fig.~\ref{fig:arch}, is organized around three components: a unified controller interface (§\ref{sec:interface}), closed-loop evaluation (§\ref{sec:harness}), and dual-metric scoring (§\ref{sec:scoring}). In the following, we describe these three platform components in more detail.

\subsection{Fairness Boundary: The Controller Interface}
\label{sec:interface}
To ensure a fair comparison across controllers, the platform requires all methods to operate within a unified and restricted decision boundary. At each control step, a controller may only act on the basis of the current system observation, either by keeping the existing allocation unchanged or by making a scheduling decision for a particular stream. To this end, the platform abstracts scheduling behavior into four action modes with stable semantics: FULL, LITE, OFFLOAD, and SKIP, corresponding respectively to full local processing, degraded local processing, remote execution, and skipping the current window. This abstraction captures the principal decision space of edge AI scheduling while avoiding ties to implementation-specific details such as model choice, input resolution, or node configuration, thereby enabling different controllers to be evaluated fairly through online decision-making under a shared task semantics.

The observation is identical for all controllers. It includes per-stream metrics (one-second p95 latency, effective fps, per-window output count, and current processing mode), per-node state (queue depth, rolling inference time, and online status), and cluster-level summaries (GPU/CPU utilization, power draw, and mode distribution). Thus, every controller, regardless of its internal decision logic, operates under the same information boundary.

The platform also fixes two interface-level rules that would otherwise affect comparison. A mode cannot be exited until it has remained active for eight control steps, and over-eager switch requests are dropped. When a controller requests OFFLOAD without specifying a target node, the platform selects the destination by minimum expected completion time (ECT), \( \mathrm{ECT} = (q + 1)\cdot\bar{t}_{\text{infer}} \), as a common default for all controllers.

\subsection{Closed-Loop Control Harness}
\label{sec:harness}

To evaluate different controllers under comparable conditions, the platform adopts a closed-loop control harness that places workload evolution and control decisions within the same execution loop. Workloads are defined in advance as scenario scripts---ordered lists of timed stream-arrival and -departure events---and replayed event by event by the platform, so that all controllers experience identical pressure trajectories; as a result, observed differences primarily reflect how controllers respond to the same scenario rather than differences in the test environment itself.

At each control step, the platform executes workload events, state observation, decision invocation, and action dispatch in a fixed order, while recording the duration of each phase separately. This design not only preserves consistency in system evolution, but also explicitly separates controller decision time from platform runtime overhead: only the duration of the decision-invocation phase is counted as scheduler overhead, while the remaining phases are reported as platform overhead. After dispatch, the platform uses compensated sleep to maintain a fixed one-second control period, and the initial warmup step is excluded from the episode summary.

\subsection{Dual-Metric SLO Scoring}
\label{sec:scoring}

A single aggregated SLO score can conflate how a controller performs overall with how it performs once its decisions have settled. To make this difference explicit, the platform is built around two primary SLO measures: raw SLO captures overall compliance across all non-SKIP windows, while steady-state SLO excludes transient windows after mode switches and reflects compliance only during stable operation. Their difference is further reported as transition tax, which makes the effect of switching explicit. For example, under heavy load, two controllers separated by 3\,pp on Raw SLO collapse to near-parity on Steady-State SLO, with the gap accounted for almost entirely by transition tax (\S\ref{sec:dual_metric}). 

Concretely, the platform evaluates each stream in each one-second window against three conditions: latency, effective frame rate, and detection quality. SKIP windows are not counted as failures, but are instead constrained by a separate episode-level budget. On this basis, the platform reports Raw SLO, Steady-state SLO, and their difference, rather than collapsing them into a single score that would implicitly impose a subjective weighting across operating phases. In parallel, the platform reports per-step scheduler and platform overheads alongside the SLO metrics, so that decision cost can be examined independently of SLO quality rather than silently absorbed into it.

\section{Evaluation}
\label{sec:eval}

This section evaluates whether the three platform components can provide the expected utility---quantified through three research questions (RQ1--RQ3). To address the three RQs, we use our benchmarking platform to compare six scheduling controllers with distinct design philosophies across five cluster configurations and two workload regimes.

\subsection{Experimental Setup}
\label{sec:setup}
\begin{table}[!b]
\centering
\footnotesize
\caption{Experimental configuration.}
\label{tab:setup}
\begin{tabular}{@{}p{0.32\columnwidth}p{0.58\columnwidth}@{}}
\toprule
\textbf{Parameter} & \textbf{Value} \\
\midrule
Coordinator       & Jetson Orin (5.3 TFLOPS) \\
GPU workers       & up to 2$\times$ Jetson Orin Nano (1.7 TFLOPS) \\
CPU workers       & up to 6$\times$ Ryzen~7, 3 throttle tiers \\
Clusters          & 2-homo, 2-het, 5-node, 8-node (LIGHT); 8-node (HEAVY) \\
Model             & YOLOv8n~\cite{jocher2023yolov8} @ 640$\times$640 \\
Video             & 720p urban traffic, 30\,fps \\
Control interval  & 1\,Hz \\
SLO latency       & p95 $\leq$ 200\,ms \\
SLO fps           & $\geq$ 10 (OFFLOAD $\geq$ 5) \\
SLO DQR           & $\geq$ 0.35 \\
Min dwell         & 8 steps \\
Scenarios         & 4 LIGHT (peak 6--8 streams); 4 HEAVY (peak 10--13) \\
Controllers       & NoOp, Least-Loaded, Heuristic, Myopic, DQN~\cite{hasselt2016ddqn}, AIF~\cite{parr2022active}; all 6 run on LIGHT and HEAVY \\
Runs per cell     & $n{=}3$ \\
Total episodes    & 424 \\
\bottomrule
\end{tabular}
\end{table}

Table~\ref{tab:setup} summarizes the experimental setup: we vary configuration and workload separately to test whether controller performance shifts across settings and under pressure (i.e. HIGH and LOW load). Because all controllers share the same interface, any observed difference in controllers' performance can be directly attributed to the controller, marginalizing the effect of the experimental environment.

DQN is pretrained per (cluster, scenario) pair with a uniform budget, so its HEAVY (high-load, 10--13 streams) results reflect in-distribution rather than transfer performance (training-budget caveat in \S\ref{sec:limits}).

\subsubsection{Evaluation Scenarios}

Each cluster is evaluated on four stress scenarios that probe distinct temporal request patterns:
\begin{itemize}
\item \textbf{Ramp-up}: active stream count rises from zero to peak over the first half of the episode, then held.
\item \textbf{Burst}: steady baseline with short 2--3 stream bursts added and removed every ${\sim}25$\,s.
\item \textbf{Steady-overload}: peak stream count applied from $t{=}0$ and held throughout.
\item \textbf{Oscillating}: stream count follows a $\sin$-like pattern with ${\sim}60$\,s period.
\end{itemize}
Each scenario runs at two regimes: \textbf{LIGHT} (peak 6--8 streams, within every cluster's sustainable GPU capacity) and \textbf{HEAVY} (peak 10--13 streams, deliberately pushing the 8-node cluster beyond local throughput).

\subsubsection{Scheduling Controllers}

We compare six controllers spanning distinct decision-making paradigms:
\begin{itemize}
\item \textbf{NoOp} (\emph{passive baseline}): never changes processing mode.

\item \textbf{Least-Loaded} (\emph{rule-based}): dispatches OFFLOAD to the node with minimum ECT; no mode switching.

\item \textbf{Heuristic} (\emph{rule-based}): thresholded state machine; promotes SKIP$\to$LITE$\to$FULL when the per-stream SLO margin is high, demotes when queue backlog is high.
\item \textbf{Myopic} (\emph{1-step greedy}): selects the $(\text{stream}, \text{mode}, \text{node})$ triple minimizing a precomputed latency--quality cost.

\item \textbf{DQN} (\emph{deep RL})~\cite{hasselt2016ddqn}: Double DQN, one seed-0 checkpoint per (cluster, scenario) with a 10-episode budget; HEAVY-workload results therefore reflect in-distribution performance, not transfer.

\item \textbf{AIF} (\emph{active inference})~\cite{parr2022active}: hierarchical generative model with hand-designed $(A, B, C, D)$ matrices and planning horizon 3.
\end{itemize}

\subsubsection{Reproducibility}

To enable third-party verification of our results, we release three classes of artifact:
\begin{itemize}
\item \textbf{Scenario scripts}: the exact workload traces (stream arrivals and departures) used in every experiment.
\item \textbf{DQN and AIF checkpoints}: the trained policies reported in Section~\ref{sec:eval}, allowing reviewers to skip the multi-hour training stage.
\item \textbf{Per-step execution histories}: the raw data from which every figure and table in this paper is regenerated.
\end{itemize}
On the hardware summarized in Table~\ref{tab:setup}, the full 424-episode sweep completes in ${\approx}70$\,hours inside an NVIDIA Container Toolkit Docker image; all artifacts are available through a public code repository\footnotemark.
\footnotetext{\url{https://anonymous.4open.science/r/cmms-benchmark-7D80}}

\subsection{Controller Performance across Load Pattern (RQ1)}
\label{sec:winner}
\begin{table}[!tb]
\centering
\small
\setlength{\tabcolsep}{4pt}
\caption{Raw SLO compliance (\%, mean $\pm 1\sigma$) across 5 cluster configurations and 6 controllers. Bold = per-cluster winner.}
\label{tab:cluster_slo}
\begin{tabular}{@{}l c c c c c@{}}
\toprule
Ctrl & 2-homo & 2-het & 5-node & 8-L & 8-H \\
\midrule
NoOp         & 57$\pm$22 & 57$\pm$22 & 56$\pm$22 & 55$\pm$22 & 19$\pm$15 \\
Least-Loaded & 84$\pm$1  & 83$\pm$2  & 84$\pm$1  & 84$\pm$2  & 44$\pm$17 \\
Myopic       & 78$\pm$3  & 78$\pm$3  & 79$\pm$4  & 79$\pm$4  & 52$\pm$14 \\
Heuristic    & 84$\pm$2  & 84$\pm$1  & 84$\pm$2  & 84$\pm$2  & \textbf{74$\pm$5}  \\
DQN          & \textbf{86$\pm$6} & \textbf{88$\pm$6} & 91$\pm$4  & 90$\pm$7  & 46$\pm$14 \\
AIF          & 79$\pm$7  & 80$\pm$6  & \textbf{93$\pm$4} & \textbf{91$\pm$5} & 71$\pm$6  \\
\bottomrule
\end{tabular}
\end{table}

\begin{figure}[!t]            
\centering                                                      
\subfloat[LIGHT ramp-up\label{fig:temporal_light}]{%
\includegraphics[width=0.95\columnwidth]{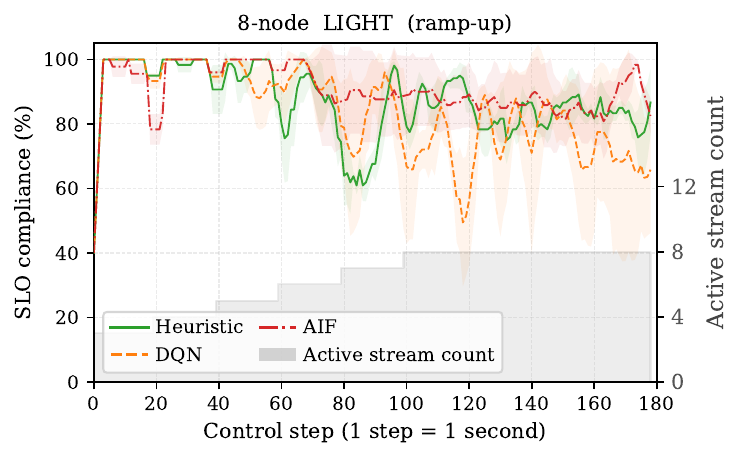}}\\                  
\vspace{-0.3em}                                                                                                                                                                  
\subfloat[HEAVY ramp-up\label{fig:temporal_heavy}]{%
\includegraphics[width=0.95\columnwidth]{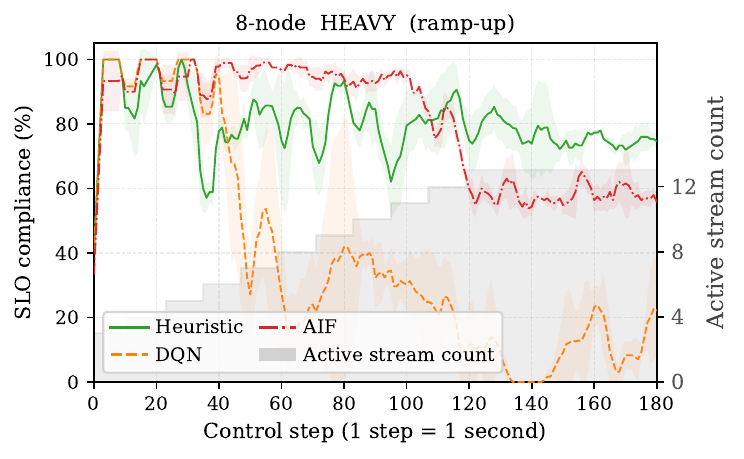}}                                                                                                       
\caption{Per-step SLO compliance on the 8-node cluster under two workload regimes. Gray shading shows active stream count; curves show per-controller SLO ($n{=}3$, 5-step moving average, $\pm1\sigma$). Each control step corresponds to one second of wall-clock time.}  
\end{figure} 

To give a fair comparison between different scheduling controllers, we evaluate controllers across increasing cluster sizes and gradually more challenging workload pattern. 
In particular, we compare per-cluster Raw SLO compliance across all 5 cluster configurations and 6 controllers (Table~\ref{tab:cluster_slo}; per-cluster winner in bold), and examine per-step SLO traces on the 8-node cluster under LIGHT (Fig.~\ref{fig:temporal_light}) and HEAVY (Fig.~\ref{fig:temporal_heavy}) ramp-up to see whether the identity of the leading controller holds across workload regimes. Figures~\ref{fig:temporal_light} and~\ref{fig:temporal_heavy}, together with Table~\ref{tab:cluster_slo}, show that no controller maintains a stable ranking across configurations. Under LIGHT, Heuristic, DQN, and AIF remain within a similar performance band, and the nominal per-cluster winners (bold) fall inside the run-to-run variance. Under HEAVY, the picture inverts sharply: moving from 8-node LIGHT to HEAVY drops DQN by 44\,pp (89.9\%\,$\to$\,45.8\%), AIF by 19\,pp, and Heuristic by only 10\,pp---the rule-based controller that was not dominant under LIGHT ends up with the highest HEAVY compliance. This rank reversal is what a single-configuration benchmark would miss.

\subsection{Controller Overhead vs.\ SLO Fulfillment (RQ2)}
\label{sec:cost_quality}

\begin{figure}[t]
\centering
\includegraphics[width=\columnwidth]{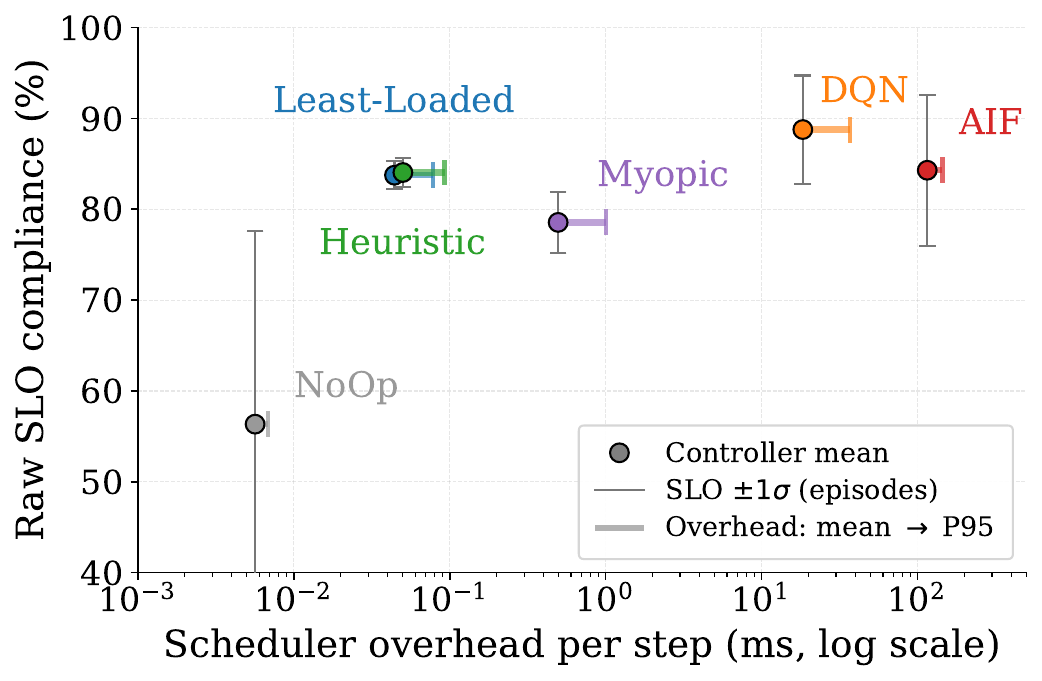}
\caption{Raw SLO versus per-step scheduler overhead (log scale) across the four LIGHT cluster configurations. Vertical bars denote SLO standard deviation across episodes; horizontal bars extending to the right of each marker span the mean-to-P95 range of per-step scheduler overhead.}  
\label{fig:cost_quality}
\end{figure}

Next, we compare the computational complexity of scheduling algorithms with their effective SLO fulfillment. Thus, we aim to construct a Pareto front that can answer for resource-constrained environments which algorithm to use.
%
Figure~\ref{fig:cost_quality} shows that decision cost and control quality are not simply aligned. Under LIGHT, per-step scheduler overhead spans several orders of magnitude across controllers, whereas Raw SLO remains within a much narrower band. In particular, AIF and Heuristic achieve nearly indistinguishable compliance (84.3\% vs.\ 84.1\%) despite a roughly 2000$\times$ gap in decision cost (115\,ms vs.\ 0.05\,ms per step), indicating that higher computational expense does not by itself imply higher control quality.

This becomes even clearer under HEAVY. As load increases, the low-cost Heuristic does not lose competitiveness; instead, it outperforms both AIF and DQN in SLO while remaining substantially cheaper per decision. Once cost and quality are reported together, the identity of the preferred controller is therefore no longer determined by compliance alone, but by the trade-off required to achieve it.

The platform can reveal this trade-off only because algorithm cost is measured separately from platform overhead. Within a given cluster, the platform-side phases remain broadly stable across controllers, with variation driven mainly by stream count rather than by the controller itself. As a result, \texttt{schedule\_ns} reflects controller cost as an algorithmic property, rather than an artifact of the harness---an aspect that prior RL-for-scheduling work~\cite{mao2017pensieve,mao2019learning} does not report alongside task-level SLO.

\subsection{Why Dual-Metric Matters (RQ3)}

\label{sec:dual_metric}
\begin{figure}[!t]
\centering
\includegraphics[width=\columnwidth]{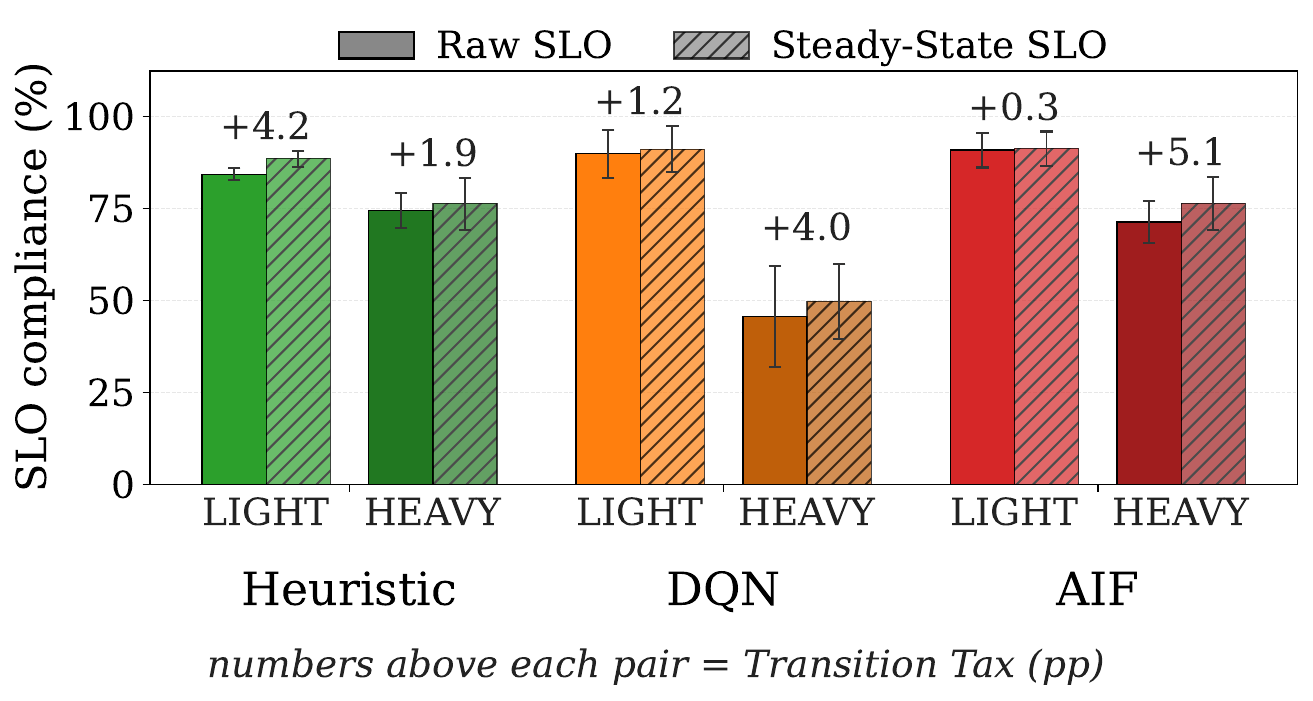}
\caption{Raw SLO (solid) and Steady-State SLO (hatched) on 8-node LIGHT (left) and HEAVY (right). Numbers above each pair report Transition Tax in percentage points.}
\label{fig:dual_metric}
\end{figure}

Lastly, we analyze the effect of separating raw from steady-state SLOs. Thus, we aim to reveal the switching costs that emerge from changing configurations, which a single (i.e., raw) score would otherwise conflate.
Figure~\ref{fig:dual_metric} shows why a single raw SLO score is not sufficient to characterize controller behavior. On 8-node HEAVY, Heuristic appears to outperform AIF by 3\,pp if comparison is based on raw SLO alone (74.5\% vs.\ 71.4\%). Once the same comparison is repeated using Steady-State SLO, however, the gap collapses to 0.14\,pp (76.3\% vs.\ 76.5\%). The difference between the two controllers therefore lies less in steady-state quality itself than in the cost incurred during switching.

This is precisely why dual-metric scoring matters. Raw SLO reflects compliance over the full execution trace, whereas Steady-State SLO isolates behavior during stable operation by excluding transient windows after mode switches. Their difference, reported as Transition Tax, makes the cost of switching explicit: AIF pays $+$5.1\,pp, nearly three times Heuristic's $+$1.9\,pp, which is where the apparent Raw-SLO gap comes from. Collapsing these two views into a single scalar would therefore conflate steady-state quality with switching cost and obscure a behaviorally important distinction between controllers. From a deployment standpoint, this distinction carries concrete trade-offs: AIF's higher switch rate is a cost under episodic workloads with long stable intervals, but a feature under rapidly shifting traffic.

\subsection{Limitations}
\label{sec:limits}

These results should be interpreted within four boundaries. First, each (cluster, controller, scenario) cell includes only $n{=}3$ runs, so 1--2\,pp gaps in the LIGHT regime should not be read as stable algorithmic rankings. Second, the HEAVY regime is evaluated only on the 8-node cluster, and its scaling behavior on smaller clusters is therefore not yet characterized.

Third, DQN uses a fixed 10-episode, single-seed training budget for each (cluster, scenario) pair. A larger budget or multi-seed training could reduce part of its HEAVY deficit, but only at substantially higher training cost. Finally, the current study is limited to one hardware family and one workload type, namely video analytics on a Jetson/Ryzen heterogeneous cluster. Generalization to other CMMS settings, such as ML serving, sensor fusion, or multi-tenant GPU scheduling, remains future work.

\section{Conclusion}
\label{sec:conclusion}

We presented a benchmark platform for Continuous Multi-Mode Scheduling on heterogeneous edge clusters and used it to study whether controller comparisons remain stable across configurations. The central finding is that they do not: controller rankings depend on cluster scale, heterogeneity, and workload regime in ways that a single-setup benchmark would hide. In particular, the apparent winner under light load does not necessarily remain competitive under heavy load, and controllers that look similar under a single aggregate score can separate once decision cost and switching effects are made explicit.

These results reinforce the role of the platform as an evaluation instrument rather than an algorithm-specific testbed. By holding the interaction boundary fixed, separating controller cost from platform overhead, and reporting both Raw and Steady-State SLO, the platform exposes differences that would otherwise be collapsed into a single number or a single scenario. The resulting picture is not that one controller is universally best, but that meaningful comparison requires multiple configurations, explicit cost accounting, and metrics that distinguish steady-state quality from transition behavior.

Our study remains limited to one hardware family and one workload type, namely video analytics on a Jetson/Ryzen heterogeneous cluster. Extending the platform to additional controller families and larger clusters, and validating it in other CMMS settings such as ML serving, sensor fusion, and multi-tenant GPU scheduling, remains future work.

\bibliographystyle{IEEEtran}
\bibliography{references}

\end{document}